\PassOptionsToPackage{unicode}{hyperref}
\PassOptionsToPackage{hyphens}{url}
\documentclass[11pt,a4paper]{article}
\usepackage{amsmath,amssymb}
\usepackage[margin=1in]{geometry}
\usepackage{pdflscape}
\usepackage{iftex}
\ifPDFTeX
  \usepackage[T1]{fontenc}
  \usepackage[utf8]{inputenc}
  \usepackage{textcomp} 
\else 
  \usepackage{unicode-math} 
  \defaultfontfeatures{Scale=MatchLowercase}
  \defaultfontfeatures[\rmfamily]{Ligatures=TeX,Scale=1}
\fi
\usepackage{lmodern}
\DeclareUnicodeCharacter{221E}{\ensuremath{\infty}}
\DeclareUnicodeCharacter{2212}{\ensuremath{-}}
\DeclareUnicodeCharacter{03B1}{\ensuremath{\alpha}}
\DeclareUnicodeCharacter{03B2}{\ensuremath{\beta}}
\DeclareUnicodeCharacter{03C6}{\ensuremath{\phi}}
\ifPDFTeX\else
\fi
\IfFileExists{upquote.sty}{\usepackage{upquote}}{}
\IfFileExists{microtype.sty}{
  \usepackage[]{microtype}
  \UseMicrotypeSet[protrusion]{basicmath} 
}{}
\makeatletter
\@ifundefined{KOMAClassName}{
  \IfFileExists{parskip.sty}{%
    \usepackage{parskip}
  }{
    \setlength{\parindent}{0pt}
    \setlength{\parskip}{6pt plus 2pt minus 1pt}}
}{
  \KOMAoptions{parskip=half}}
\makeatother
\usepackage{xcolor}
\usepackage{longtable,booktabs,array}
\usepackage{multirow}
\usepackage{calc} 
\usepackage{etoolbox}
\makeatletter
\patchcmd\longtable{\par}{\if@noskipsec\mbox{}\fi\par}{}{}
\makeatother
\IfFileExists{footnotehyper.sty}{\usepackage{footnotehyper}}{\usepackage{footnote}}
\makesavenoteenv{longtable}
\usepackage{graphicx}
\makeatletter
\def\maxwidth{\ifdim\Gin@nat@width>\linewidth\linewidth\else\Gin@nat@width\fi}
\def\maxheight{\ifdim\Gin@nat@height>\textheight\textheight\else\Gin@nat@height\fi}
\makeatother
\setkeys{Gin}{width=\maxwidth,height=\maxheight,keepaspectratio}
\makeatletter
\def\fps@figure{htbp}
\makeatother
\ifLuaTeX
  \usepackage{luacolor}
  \usepackage[soul]{lua-ul}
\else
  \IfFileExists{soul.sty}{\usepackage{soul}}{}
\fi
\ifLuaTeX
  \usepackage{selnolig}  
\fi
\usepackage{bookmark}
\IfFileExists{xurl.sty}{\usepackage{xurl}}{} 
\hypersetup{
  pdftitle={Study of flux redistribution and shadowing effect in rod worth measurements with the rod drop experiment},
  hidelinks,
  pdfcreator={LaTeX reconstructed from an original author Word manuscript},
  pdfsubject={Author manuscript; published DOI 10.1016/j.nucengdes.2023.112672},
  pdfkeywords={control rod worth measurement,rod-drop experiment,neutron flux redistribution,control rod shadowing effect,reactivity measurement and calibration,inverse point kinetics,neutron detector response,MCNP Monte Carlo neutron transport,source jerk method,IPEN/MB-01 research reactor,asymptotic period method}}

\title{Study of flux redistribution and shadowing effect in rod worth measurements with the rod drop experiment}
\author{}
\date{}

\begin{document}

\maketitle
\noindent\fbox{\begin{minipage}{0.95\linewidth}
\small\textbf{Author manuscript -- publication record.} \textit{Author Word version from October 2023; compare with the subsequently accepted R1 manuscript.}\\
The definitive peer-reviewed paper appeared in \emph{Nuclear Engineering and Design 415 (2023), 112672}.\\
Original DOI: \href{https://doi.org/10.1016/j.nucengdes.2023.112672}{10.1016/j.nucengdes.2023.112672}. \textbf{Please cite the published article as the version of record.}
\end{minipage}}\par\medskip

\begin{longtable}[]{@{}
  >{\raggedright\arraybackslash}p{(\columnwidth - 0\tabcolsep) * \real{1.0000}}@{}}
\toprule\noalign{}
\begin{minipage}[b]{\linewidth}\raggedright
Giovanni L. Stefani\textsuperscript{1}, Guilherme E. S. Carvalho\textsuperscript{1}, Daniel A. P. Palma\textsuperscript{2}, Marcelo V. Silva¹, Cesar Raitz³

\textsuperscript{1}Programa de Engenharia Nuclear da COPPE/UFRJ

Universidade Federal do Rio de Janeiro

Av. Horácio Macedo, 2030, Cidade Universitária

21941-941Rio de janeiro- RJ - Brazil,

giovanni.laranjo@coppe.ufrj.br

\textsuperscript{2}Comissão Nacional de Energia Nuclear CNEN:

Rua Gal Severiano, 90,

22290-901 Rio de Janeiro - RJ -- Brazil

³ Instituto de Engenharia Nuclear

R. Hélio de Almeida, 75, Cidade Universitária

21941-614 Rio de Janeiro -- Rj - Brazil
\end{minipage} \\
\midrule\noalign{}
\endhead
\bottomrule\noalign{}
\endlastfoot
\end{longtable}

\section{ABSTRACT}\label{abstract}

The main goal  of this study is to quantify the flux redistribution in neutron detectors and the shadowing effect between control rods by proposing an instrumental calibration factor for measurement, based on computer simulations. The MCNP code was used to simulate the rod drop experiment in the IPEN/MB-01 research reactor. A study of the reactivity behavior was carried out during the most transient period of the stick falling experiment (t \textless{} 10 s), demonstrating great agreement between the simulation and the experiment carried out using the source jerk method. A methodology was developed to estimate the shadowing effect, flux redistribution and propose a correction of the results obtained (calibration) in the simulations using the MCNP code and the rod drop experiment respectively. The maximum deviations between the reference value obtained from measuring the control banks via the stable period method and the simulated values were 0.5\% and 11.3\%, respectively.

\textbf{Keywords:} control rod worth measurement; rod-drop experiment; neutron flux redistribution; control rod shadowing effect; reactivity measurement and calibration; inverse point kinetics; neutron detector response; MCNP Monte Carlo neutron transport; source jerk method; IPEN/MB-01 research reactor; asymptotic period method.

\begin{enumerate}
\def\labelenumi{\arabic{enumi}.}
\item
  \textbf{INTRODUCTION}
\end{enumerate}

After a core reload or other significant changes to the reactor core, physical start-up tests on nuclear reactors are required to ensure that the operating characteristics are within their specifications. One of the most important phases of the physical start-up tests is to experimentally obtain the control and safety rod worth measurement. The measures obtained in the tests are compared with the calculated ones, which are considered as references. Regardless of the operating purposes (which may be electric power generation, radioisotope production and research, or the reactivity control mode) all nuclear reactors have, as part of their design, control sets and/or safety rods, whose reactivity must be well known and, mainly in power reactors, must be verified by means of direct measurements (Kloosterman, Rugama, and Szieberth, 2002).

Some works enshrined in the literature that used techniques to measure the rod worth such as the ``neutron source multiplication method'' (Endo et. al., 2011; Pyeon et al., 2017), the ``neutron oscillation method'' bar'' (Trkov et. al., 1995), ``Source Jerk'' (Ryves and Scott, 1962; Roy et. al., 2017), ``boron dilution (Merljak and Trkov, 2018; Holt et. al., 2016 ) , control rod insertion (Kaviani and Mirvakili, 2020) which, due to its importance, is the subject of several scientific and technological studies in reactor physics and nuclear engineering (Moreira and Lee, 1984; Gauthier et. al., 2017; Varvayanni et. al., 2009; Fadaei and Setayeshi, 2009; Mele et al., 2017; Thilagam et. al., 2021; Sembiring et. al., 2021). The rod drop is a well-known technique with the advantage of being quick and simple to perform when compared to other techniques such as ``boron dilution'', for instance. The idea of the ``rod drop'' technique is to start acting from the critical state, or as close as possible to it, and then drop the control rod in relation to the position in which the integral reactivity will be measured. The neutron flux is monitored and used to obtain the reactivity through the inverse point kinetic equations. In addition, due to the large insertion of negative reactivity caused by the control banks, the neutron flux has its distribution shape changed, which causes the shadowing effect on the neutron detectors, especially those that are around the position where the control bar was inserted. .

The objective of this paper is to quantify the shadowing effect due more than one control rod presence, quantify the redistribution of the neutron flux in neutron detectors during the insertion of control rods and to propose a correction factor, based on computational simulations, for reliable measurements of reactivity. This paper is structured as follows: section 2 will present the methodology used, based on the asymptotic period (Son et. al., 2017) and inverse kinetics (Duderstadt, 1976) methods. In section 3, the IPEN/MB-01 reactor, the installation considered in the present paper, is described in detail in order to support the simulations carried out. In section 4 the results obtained are reported and in section 5 the conclusions reached and some discussions about the shading phenomenon are presented.

\begin{enumerate}
\def\labelenumi{\arabic{enumi}.}
\setcounter{enumi}{1}
\item
  METHODOLOGY
\end{enumerate}

2.1. Asymptotic period method

The asymptotic period method is the most commonly used method (Asuku et. al., 2015; Nelson and Doane, 1990) for calibrating control rods. The procedure consists of the correlation that exists between the stability of the reactor, the time that the reactor takes to change its power by a factor e, and the reactivity of the reactor using the kinetic relations derived from the diffusion equation:

\(\frac{dn}{dt} = \frac{\rho - \beta}{\Lambda}n + \sum_{i = 1}^{6}\lambda_{i}C_{i}\) (1)

\(\frac{dC_{i}}{dt} = \frac{\beta_{i}}{\Lambda}n - \lambda_{i}C_{i}\) , (2)

where n is the neutron density, \(\rho\) a reactivity, \(\beta\) the fraction of delayed neutrons, \(\beta_{i}\) the fraction of neutrons in the group i\textsuperscript{th}, \(\Lambda\) the generation time of prompt neutrons, \(\lambda_{i}\) the decay constant of the group i\textsuperscript{th} e \(C_{i}\) the concentration of precursors group i\textsuperscript{th}. The solution to this system of differential equations will be:

\(n(t) = n_{0}e^{wt}\), (3)

being that,

\(C_{i}(t) = C_{i0}e^{wt},\) (4)

where w is an undefined parameter with inverse time unit.

Substituting Eq. (3) and Eq. (4) in Eq. (2), we have that:

\(wC_{i0}e^{wt} = - \lambda_{i}C_{i0}e^{wt} + \frac{\beta_{i}}{\Lambda}n_{0}e^{wt}\) , (5)

or,

\(C_{i0} = \frac{\beta_{i}}{(\lambda_{i}\  + \ w)\Lambda}n_{0}.\) (6)

Thus, is well stablished that the concentrations of the delayed neutron precursors are proportional to the neutron density. Thus, substituting Eqs. (3), Eq. (4) and Eq. (6) in Eq. (1), we get the following expression:

\(w = \left( \frac{\rho - \beta}{\Lambda} \right) + \sum_{i = 1}^{6}\lambda_{i}\left\lbrack \frac{\beta_{i}}{(\lambda_{i}\  + \ w)\Lambda} \right\rbrack\ ,\) (7)

where, \(\Lambda \equiv \frac{l}{k}\) is the mean generation time between birth and subsequent neutron absorption. Furthermore, considering the definition of \(\beta \equiv \sum_{i = 1}^{6}\beta_{i}\) as the fraction of delayed neutrons, Eq. (7) can be rewritten as follows:

\(w\Lambda = \rho - \sum_{i = 1}^{6}\left\lbrack \frac{w\beta_{i}}{(\lambda_{i}\  + \ w)} \right\rbrack\ \) . (8)

The reactivity is expressed approximately by:

\(\rho(t) \cong \frac{k(t) - 1}{k(t)}\) . (9)

Isolating the term reactivity, we finally have:

\(\rho = w\Lambda + \sum_{i = 1}^{6}\frac{w\beta_{i}}{w + \lambda_{i}}\) . (10)

This expression is known as the Inhour Equation (Duderstadt and Hamilton, 1976; Marguet, 2018).

The solution will be:

\(n(t) = A_{0}e^{w_{0}t} + A_{1}e^{w_{1}t} + \ldots + A_{6}e^{w_{6}t}\) , (11)

where w\textsubscript{i} are the roots of the reactivity equation. If the reactivity is positive, we will find a positive root \(w_{0}\) and 6 negative ones. Thus, after a short time the neutron density reduces to:

\(n(t) = A_{0}e^{w_{0}t}\) . (12)

By definition, the stable period (T\textsubscript{p}) of the reactor will be:

\(T_{p} = \frac{1}{w_{0}}\) , (13)

and also that Eq. (10) will be writing by:

\(\rho = \frac{\Lambda}{T_{p}} + \sum_{i = 1}^{6}\frac{\beta_{i}}{1 + T_{p}\lambda_{i}}\) . (14)

Defining the period \(T_{p}\) as the time required for the reactor power to increase by a factor e = 2.7182, and doubling time \ensuremath{\delta} as the time required for the reactor power to increase by a factor 2.

The experiment for this method consists of criticalizing the reactor with the control rod and then partially removing this control rod, so the reactor will become supercritical. Then the period is measured with the aid of a stopwatch to obtain the corresponding reactivity. Then the other control rod is inserted so that the reactor returns to criticality.

2.2. Inverse kinetics method

The inverse kinetics method consists of a real-time and continuous measurement of the reactivity, which is derived from the reactor kinetics equation (1):

\(\rho(t) = \beta + \frac{\Lambda}{n}\frac{dn}{dt} - \frac{1}{n}\sum_{i = 1}^{6}\lambda_{i}C_{i}\) , (15)

while the precursor equation is obtained by integrating equation (2):

\(C_{i}(t) = C_{i}(0)e^{- \lambda_{i}t} + \frac{\beta_{i}}{\Lambda}\int_{0}^{t}n(\tau)e^{- \lambda_{i}\tau}d\tau\) . (16)

Assuming the critical reactor at time t = 0, from (2) it is possible to write that:

\(C_{i}(0) = \frac{\beta_{i}}{\lambda_{i}\Lambda}n(0)\) (17)

and that the reactivity is written by the following equation:

\(\rho(t) = \beta + \frac{\Lambda}{n}\frac{dn}{dt} - \frac{n(0)}{n}\sum_{i = 1}^{6}\beta_{i}e^{- \lambda_{i}t} - \frac{1}{n}\sum_{i = 1}^{6}\lambda_{i}\beta_{i}e^{- \lambda_{i}t}\int_{0}^{t}n(\tau)e^{- \lambda_{i}\tau}d\tau\) (18)

Thus, it is possible associate the proportionality of the current measured in the detector with the neutron flux in the reactor.

2.3. The rod drop method

The rod drop method consists of an experiment in which the reactor is operated at a constant power and then a scram is realized by dropping a set of control rods by gravitational force. The neutron flux measurement must be done after the decay of the higher modes, resulting only in the fundamental one. There is a time gap between ten and sixty seconds before the decay of the delayed neutrons.

This method depends on the following assumptions (Buoni, 1963):

\begin{enumerate}
\def\labelenumi{\arabic{enumi})}
\item
  Before the control rods drop, the precursors of delayed neutrons must be in balance;
\item
  The kinetics equation for the precursors of delayed neutrons does not depends on the space;
\item
  The multiplication source is given by \(M = \frac{1}{1 - k}\);
\item
  After dropping the set of control rods, the generation of delayed neutrons can be neglected;
\item
  The detector presents the same neutrons population before and immediately after the control rods drop in a moment ``t''.
\end{enumerate}

The first assumption, in which the delayed neutrons must be in balance, can be attended by operating the reactor in a critical state for around ten minutes before the drop of the set of control rods.

The second assumption is very accurate for the purposes of verifying the reactivity with the rod drop method.

The third assumption is not verified. The result of the neutron flux given by \(\varphi = MS\), with \emph{S} being the neutron flux only due to the source, is an auto-stationary asymptotic state. However, the period after the drop of the control rods, in which the behavior of the source of delayed neutrons will be considered, must be sufficiently slow to be acceptable, i.e. low transient variation of reactivity over time.

The fourth assumption is valid if, after one second, the neutron flux decays at least two decades.

The fifth assumption implies that the neutron flux through the detectors is proportional to the neutron flux inside the reactor core and the fall of the control rods does not change its spatial distribution, only its amplitude. This assumption has never been verified and, in this word, we demonstrate that the rod drop redistributes the neutron flux through the detectors, which changes the reactivity measurement. To correct this redistribution, a shape factor that will consider the redistribution of the neutron flux will be proposed. .

To test the rod drop method, an experiment was performed in the IPEN/MB-01 reactor in a 28x26 configuration. The scram was achieved by dropping the reactor\textquotesingle s set A of control rods (Figure 2). The neutron flux was measured using the four detectors placed around the core and their values were transformed in reactivity measure by using an instrument called reactimeter (Moreira, 1986).

3. THE IPEN/MB-01 REACTOR

The IPEN/MB-01 reactor is known as the zero power type and is designed to operate with a maximum power of less than 100 watts. These types of reactors are a resource that allows researchers to study not only theoretical calculations, but also experimental measurements. In the present paper, the IPEN/MB-01 reactor was chosen to carry out the experiments due to the relevance of being an installation that allows the simulation of nuclear characteristics of large reactors on a reduced scale.

In this section, the description of the reactor used both in the experiments and for the computational modeling will be presented. Table 1 illustrates the geometric data of the reactor, and in figure 1 it is possible to visualize a radial section of the fuel rod. The modeling information were based on references (Santos et. al, 2013; Santos et. al., 2008; Santos et al, 2006).

\textbf{Table 1:} Definitions of the geometry, material and density used in IPEN/MB-01. Power set in 100 W.

\begin{longtable}[]{@{}
  >{\raggedright\arraybackslash}p{(\columnwidth - 6\tabcolsep) * \real{0.3837}}
  >{\raggedright\arraybackslash}p{(\columnwidth - 6\tabcolsep) * \real{0.2041}}
  >{\raggedright\arraybackslash}p{(\columnwidth - 6\tabcolsep) * \real{0.2041}}
  >{\raggedright\arraybackslash}p{(\columnwidth - 6\tabcolsep) * \real{0.2081}}@{}}
\toprule\noalign{}
\begin{minipage}[b]{\linewidth}\raggedright
\textbf{Reactor component}
\end{minipage} & \begin{minipage}[b]{\linewidth}\raggedright
\textbf{Length (cm)}
\end{minipage} & \begin{minipage}[b]{\linewidth}\raggedright
\textbf{Material}
\end{minipage} & \begin{minipage}[b]{\linewidth}\raggedright
\textbf{Atomic Density (atom/cm Barn)}
\end{minipage} \\
\begin{minipage}[b]{\linewidth}\raggedright
Rod Pitch
\end{minipage} & \begin{minipage}[b]{\linewidth}\raggedright
1.50000
\end{minipage} & \begin{minipage}[b]{\linewidth}\raggedright
Water
\end{minipage} & \begin{minipage}[b]{\linewidth}\raggedright
1.00104E-01
\end{minipage} \\
\begin{minipage}[b]{\linewidth}\raggedright
Pellet Radius
\end{minipage} & \begin{minipage}[b]{\linewidth}\raggedright
0.42447
\end{minipage} & \begin{minipage}[b]{\linewidth}\raggedright
UO2 -- 4.3 w/o
\end{minipage} & \begin{minipage}[b]{\linewidth}\raggedright
6.819010E-02
\end{minipage} \\
\begin{minipage}[b]{\linewidth}\raggedright
Inner clad radius
\end{minipage} & \begin{minipage}[b]{\linewidth}\raggedright
0.42873
\end{minipage} & \begin{minipage}[b]{\linewidth}\raggedright
He
\end{minipage} & \begin{minipage}[b]{\linewidth}\raggedright
1.504520E-05
\end{minipage} \\
\begin{minipage}[b]{\linewidth}\raggedright
Outer clad radius
\end{minipage} & \begin{minipage}[b]{\linewidth}\raggedright
0.49037
\end{minipage} & \begin{minipage}[b]{\linewidth}\raggedright
Stainless Steel
\end{minipage} & \begin{minipage}[b]{\linewidth}\raggedright
8.657177E-02
\end{minipage} \\
\begin{minipage}[b]{\linewidth}\raggedright
Fuel height
\end{minipage} & \begin{minipage}[b]{\linewidth}\raggedright
54.84000
\end{minipage} & \begin{minipage}[b]{\linewidth}\raggedright
\end{minipage} & \begin{minipage}[b]{\linewidth}\raggedright
\end{minipage} \\
\begin{minipage}[b]{\linewidth}\raggedright
Inner Control rod radius (Ag-In-Cd)
\end{minipage} & \begin{minipage}[b]{\linewidth}\raggedright
0.41600
\end{minipage} & \begin{minipage}[b]{\linewidth}\raggedright
Ag/In/Cd
\end{minipage} & \begin{minipage}[b]{\linewidth}\raggedright
6.131741E-02
\end{minipage} \\
\begin{minipage}[b]{\linewidth}\raggedright
Inner clad radius (control rod)
\end{minipage} & \begin{minipage}[b]{\linewidth}\raggedright
0.42873
\end{minipage} & \begin{minipage}[b]{\linewidth}\raggedright
He
\end{minipage} & \begin{minipage}[b]{\linewidth}\raggedright
1.504520E-05
\end{minipage} \\
\begin{minipage}[b]{\linewidth}\raggedright
Outer clad radius (Control rod)
\end{minipage} & \begin{minipage}[b]{\linewidth}\raggedright
0.49037
\end{minipage} & \begin{minipage}[b]{\linewidth}\raggedright
Stainless Steel
\end{minipage} & \begin{minipage}[b]{\linewidth}\raggedright
8.657177E-02
\end{minipage} \\
\begin{minipage}[b]{\linewidth}\raggedright
Control rod height
\end{minipage} & \begin{minipage}[b]{\linewidth}\raggedright
70.34000
\end{minipage} & \begin{minipage}[b]{\linewidth}\raggedright
\end{minipage} & \begin{minipage}[b]{\linewidth}\raggedright
\end{minipage} \\
\midrule\noalign{}
\endhead
\bottomrule\noalign{}
\endlastfoot
\end{longtable}

\includegraphics[width=3.77698in,height=2.4093in]{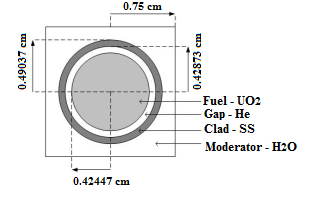}

\textbf{Figure 1:} Fuel rod (radial view).

The composition of the materials used in the reactor core is described in Table 2, as well as its temperature. The IPEN/MB-01 reactor is a zero-power type, thus it keeps an isothermal temperature of 300K in its core.

\textbf{Table 2:} Material composition.

\begin{longtable}[]{@{}
  >{\raggedright\arraybackslash}p{(\columnwidth - 6\tabcolsep) * \real{0.2847}}
  >{\raggedright\arraybackslash}p{(\columnwidth - 6\tabcolsep) * \real{0.1696}}
  >{\raggedright\arraybackslash}p{(\columnwidth - 6\tabcolsep) * \real{0.2899}}
  >{\raggedright\arraybackslash}p{(\columnwidth - 6\tabcolsep) * \real{0.2558}}@{}}
\toprule\noalign{}
\begin{minipage}[b]{\linewidth}\raggedright
\textbf{Component}
\end{minipage} & \begin{minipage}[b]{\linewidth}\raggedright
\textbf{Material}
\end{minipage} & \begin{minipage}[b]{\linewidth}\raggedright
\textbf{Atomic Density (atom/cm Barn)}
\end{minipage} & \begin{minipage}[b]{\linewidth}\raggedright
\textbf{Temperature}
\end{minipage} \\
\multirow{3}{=}{\begin{minipage}[b]{\linewidth}\raggedright
Fuel
\end{minipage}} & \begin{minipage}[b]{\linewidth}\raggedright
U-235
\end{minipage} & \begin{minipage}[b]{\linewidth}\raggedright
9.99240E-04
\end{minipage} & \multirow{15}{=}{\begin{minipage}[b]{\linewidth}\raggedright
300 K
\end{minipage}} \\
& \begin{minipage}[b]{\linewidth}\raggedright
U-238
\end{minipage} & \begin{minipage}[b]{\linewidth}\raggedright
2.1694E-02
\end{minipage} \\
& \begin{minipage}[b]{\linewidth}\raggedright
O-16
\end{minipage} & \begin{minipage}[b]{\linewidth}\raggedright
4.54890e-02
\end{minipage} \\
\multirow{5}{=}{\begin{minipage}[b]{\linewidth}\raggedright
Clad
\end{minipage}} & \begin{minipage}[b]{\linewidth}\raggedright
Fe
\end{minipage} & \begin{minipage}[b]{\linewidth}\raggedright
5.67582E-02
\end{minipage} \\
& \begin{minipage}[b]{\linewidth}\raggedright
Ni
\end{minipage} & \begin{minipage}[b]{\linewidth}\raggedright
8.64435E-03
\end{minipage} \\
& \begin{minipage}[b]{\linewidth}\raggedright
Cr
\end{minipage} & \begin{minipage}[b]{\linewidth}\raggedright
1.72649E-02
\end{minipage} \\
& \begin{minipage}[b]{\linewidth}\raggedright
Mn
\end{minipage} & \begin{minipage}[b]{\linewidth}\raggedright
1.59898E-03
\end{minipage} \\
& \begin{minipage}[b]{\linewidth}\raggedright
Si
\end{minipage} & \begin{minipage}[b]{\linewidth}\raggedright
3.34513E-04
\end{minipage} \\
\multirow{5}{=}{\begin{minipage}[b]{\linewidth}\raggedright
Control rods
\end{minipage}} & \begin{minipage}[b]{\linewidth}\raggedright
Ag-107
\end{minipage} & \begin{minipage}[b]{\linewidth}\raggedright
2.35462E-02
\end{minipage} \\
& \begin{minipage}[b]{\linewidth}\raggedright
Ag-109
\end{minipage} & \begin{minipage}[b]{\linewidth}\raggedright
2.18835E-02
\end{minipage} \\
& \begin{minipage}[b]{\linewidth}\raggedright
In-113
\end{minipage} & \begin{minipage}[b]{\linewidth}\raggedright
3.42506E-04
\end{minipage} \\
& \begin{minipage}[b]{\linewidth}\raggedright
In-115
\end{minipage} & \begin{minipage}[b]{\linewidth}\raggedright
7.6599E-03
\end{minipage} \\
& \begin{minipage}[b]{\linewidth}\raggedright
Cd
\end{minipage} & \begin{minipage}[b]{\linewidth}\raggedright
2.72492E-03
\end{minipage} \\
\begin{minipage}[b]{\linewidth}\raggedright
Water
\end{minipage} & \begin{minipage}[b]{\linewidth}\raggedright
O-16
\end{minipage} & \begin{minipage}[b]{\linewidth}\raggedright
3.33680E-02
\end{minipage} \\
\begin{minipage}[b]{\linewidth}\raggedright
\end{minipage} & \begin{minipage}[b]{\linewidth}\raggedright
H-1
\end{minipage} & \begin{minipage}[b]{\linewidth}\raggedright
6.67360E-02
\end{minipage} \\
\midrule\noalign{}
\endhead
\bottomrule\noalign{}
\endlastfoot
\end{longtable}

The reactor has a rectangular configuration of 28x26 rods, as shown in figure 2. In this configuration, \textbf{A} and \textbf{B} are the control rods and \textbf{S} are the safety rods, the latter of which will not be used in the computational modeling because they are not necessary for the experiment.

\includegraphics[width=4.04167in,height=2.70833in]{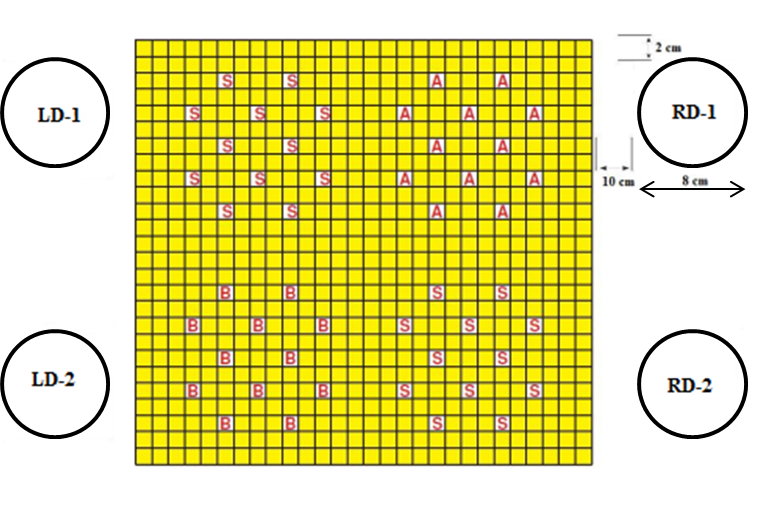}

\textbf{Figure 2:} Reactor IPEN/MB-01 -- 28x26 configuration.

The neutron detectors are positioned in four positions: LD-1, LD-2, RD-1 and RD-2, as shown in Figure 2. The detectors have a circular shape with a radius of 4 cm and a length approximately equal to the active region (54.84 cm), i.e. the region containing uranium axially. In this study, the detectors were simplified to a control volume, made of water, with the described geometric characteristics.

\textbf{Table 3:} Insertion of control and safety rods in a critical state of the IPEN/MB-01 reactor with a 28x26 configuration.

\begin{longtable}[]{@{}
  >{\raggedright\arraybackslash}p{(\columnwidth - 6\tabcolsep) * \real{0.2576}}
  >{\raggedright\arraybackslash}p{(\columnwidth - 6\tabcolsep) * \real{0.2536}}
  >{\raggedright\arraybackslash}p{(\columnwidth - 6\tabcolsep) * \real{0.2746}}
  >{\raggedright\arraybackslash}p{(\columnwidth - 6\tabcolsep) * \real{0.2143}}@{}}
\toprule\noalign{}
\begin{minipage}[b]{\linewidth}\raggedright
\end{minipage} & \multicolumn{3}{>{\raggedright\arraybackslash}p{(\columnwidth - 6\tabcolsep) * \real{0.7424} + 4\tabcolsep}@{}}{%
\begin{minipage}[b]{\linewidth}\raggedright
\textbf{Insertion in the active region (\%)}
\end{minipage}} \\
\begin{minipage}[b]{\linewidth}\raggedright
Configuration
\end{minipage} & \begin{minipage}[b]{\linewidth}\raggedright
Safety (S)
\end{minipage} & \begin{minipage}[b]{\linewidth}\raggedright
Control (A)
\end{minipage} & \begin{minipage}[b]{\linewidth}\raggedright
Control (B)
\end{minipage} \\
\begin{minipage}[b]{\linewidth}\raggedright
1
\end{minipage} & \begin{minipage}[b]{\linewidth}\raggedright
0
\end{minipage} & \begin{minipage}[b]{\linewidth}\raggedright
0
\end{minipage} & \begin{minipage}[b]{\linewidth}\raggedright
68
\end{minipage} \\
\begin{minipage}[b]{\linewidth}\raggedright
2
\end{minipage} & \begin{minipage}[b]{\linewidth}\raggedright
0
\end{minipage} & \begin{minipage}[b]{\linewidth}\raggedright
42
\end{minipage} & \begin{minipage}[b]{\linewidth}\raggedright
42
\end{minipage} \\
\midrule\noalign{}
\endhead
\bottomrule\noalign{}
\endlastfoot
\end{longtable}

The ENDF/B-VII.0 nuclear data library, which contains more detailed and specific information on nuclear reaction cross sections for a wide variety of nuclides, was used in the reactor modeling.

The kinetic parameters used in the calculation of the reactimeter, which is based on the inverse kinetic equation, can be seen in Table 4 below (Pinto, 2012):

Table 4: Kinetic parameters used in reactivity measurements with neutron detectors.

\begin{longtable}[]{@{}
  >{\raggedright\arraybackslash}p{(\columnwidth - 6\tabcolsep) * \real{0.2500}}
  >{\raggedright\arraybackslash}p{(\columnwidth - 6\tabcolsep) * \real{0.2500}}
  >{\raggedright\arraybackslash}p{(\columnwidth - 6\tabcolsep) * \real{0.2500}}
  >{\raggedright\arraybackslash}p{(\columnwidth - 6\tabcolsep) * \real{0.2501}}@{}}
\toprule\noalign{}
\multicolumn{2}{@{}>{\raggedright\arraybackslash}p{(\columnwidth - 6\tabcolsep) * \real{0.4999} + 2\tabcolsep}}{%
\begin{minipage}[b]{\linewidth}\raggedright
Delayed neutron precursor group decay constant (s\textsuperscript{-1})
\end{minipage}} & \multicolumn{2}{>{\raggedright\arraybackslash}p{(\columnwidth - 6\tabcolsep) * \real{0.5001} + 2\tabcolsep}@{}}{%
\begin{minipage}[b]{\linewidth}\raggedright
Delayed Neutron Precursor Concentration Fraction
\end{minipage}} \\
\begin{minipage}[b]{\linewidth}\raggedright
\includegraphics{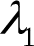}
\end{minipage} & \begin{minipage}[b]{\linewidth}\raggedright
0.012800
\end{minipage} & \begin{minipage}[b]{\linewidth}\raggedright
\includegraphics{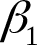}
\end{minipage} & \begin{minipage}[b]{\linewidth}\raggedright
2.07000x10\textsuperscript{-4}
\end{minipage} \\
\begin{minipage}[b]{\linewidth}\raggedright
\includegraphics{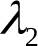}
\end{minipage} & \begin{minipage}[b]{\linewidth}\raggedright
0.031700
\end{minipage} & \begin{minipage}[b]{\linewidth}\raggedright
\includegraphics{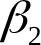}
\end{minipage} & \begin{minipage}[b]{\linewidth}\raggedright
1.29300x10\textsuperscript{-3}
\end{minipage} \\
\begin{minipage}[b]{\linewidth}\raggedright
\includegraphics{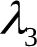}
\end{minipage} & \begin{minipage}[b]{\linewidth}\raggedright
0.120900
\end{minipage} & \begin{minipage}[b]{\linewidth}\raggedright
\includegraphics{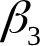}
\end{minipage} & \begin{minipage}[b]{\linewidth}\raggedright
1.17600x10\textsuperscript{-3}
\end{minipage} \\
\begin{minipage}[b]{\linewidth}\raggedright
\includegraphics{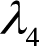}
\end{minipage} & \begin{minipage}[b]{\linewidth}\raggedright
0.321200
\end{minipage} & \begin{minipage}[b]{\linewidth}\raggedright
\includegraphics{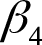}
\end{minipage} & \begin{minipage}[b]{\linewidth}\raggedright
2.53700x10\textsuperscript{-3}
\end{minipage} \\
\begin{minipage}[b]{\linewidth}\raggedright
\includegraphics{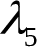}
\end{minipage} & \begin{minipage}[b]{\linewidth}\raggedright
1.400500
\end{minipage} & \begin{minipage}[b]{\linewidth}\raggedright
\includegraphics{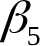}
\end{minipage} & \begin{minipage}[b]{\linewidth}\raggedright
9.37000x10\textsuperscript{-4}
\end{minipage} \\
\begin{minipage}[b]{\linewidth}\raggedright
\includegraphics{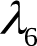}
\end{minipage} & \begin{minipage}[b]{\linewidth}\raggedright
3.878100
\end{minipage} & \begin{minipage}[b]{\linewidth}\raggedright
\includegraphics{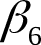}
\end{minipage} & \begin{minipage}[b]{\linewidth}\raggedright
2.31000x10\textsuperscript{-4}
\end{minipage} \\
\begin{minipage}[b]{\linewidth}\raggedright
\end{minipage} & \multicolumn{2}{>{\raggedright\arraybackslash}p{(\columnwidth - 6\tabcolsep) * \real{0.4999} + 2\tabcolsep}}{%
\begin{minipage}[b]{\linewidth}\raggedright
Prompt neutron lifetime(s):
\end{minipage}} & \begin{minipage}[b]{\linewidth}\raggedright
\end{minipage} \\
\begin{minipage}[b]{\linewidth}\raggedright
\end{minipage} & \begin{minipage}[b]{\linewidth}\raggedright
\ensuremath{\Lambda}
\end{minipage} & \begin{minipage}[b]{\linewidth}\raggedright
17.00x10\textsuperscript{-6}
\end{minipage} & \begin{minipage}[b]{\linewidth}\raggedright
\end{minipage} \\
\midrule\noalign{}
\endhead
\bottomrule\noalign{}
\endlastfoot
\end{longtable}

The data acquisition time by the reactimeter was 1.02426 s and with initial neutron flux of 1.0541x10\textsuperscript{-11} n/cm²s and the experiments carried out at room temperature \textasciitilde27 \ensuremath{^\circ}C.

4. RESULTS

\begin{enumerate}
\def\labelenumi{\arabic{enumi}.}
\item
  MCNP Modeling Validation
\end{enumerate}

To validate the modeling, let start from the evaluation of 5 known measures, namely:

\begin{itemize}
\item
  Reactivity excess of the configuration 28x26;
\item
  Fraction of delayed neutrons;
\item
  Criticalization of the 28x26 configuration with a control bank (32\% removed -- configuration 1);
\item
  Criticalization of the 28x26 configuration with two control bank (58\% removed -- configuration 2);
\item
  Control bank calibration measurement experiment.
\end{itemize}

For the first three measures, we have their comparison of values given in table 5 with the calculations via MCNP code:

Table 5: Validation of IPEN/MB-01 reactor measurements.

\begin{longtable}[]{@{}
  >{\raggedright\arraybackslash}p{(\columnwidth - 4\tabcolsep) * \real{0.3330}}
  >{\raggedright\arraybackslash}p{(\columnwidth - 4\tabcolsep) * \real{0.2725}}
  >{\raggedright\arraybackslash}p{(\columnwidth - 4\tabcolsep) * \real{0.3945}}@{}}
\toprule\noalign{}
\multirow{2}{=}{\begin{minipage}[b]{\linewidth}\raggedright
\end{minipage}} & \multicolumn{2}{>{\raggedright\arraybackslash}p{(\columnwidth - 4\tabcolsep) * \real{0.6670} + 2\tabcolsep}@{}}{%
\begin{minipage}[b]{\linewidth}\raggedright
\textbf{Excess reactivity (pcm)}
\end{minipage}} \\
& \begin{minipage}[b]{\linewidth}\raggedright
\textbf{Calculated}
\end{minipage} & \begin{minipage}[b]{\linewidth}\raggedright
\textbf{Reference}
\end{minipage} \\
\begin{minipage}[b]{\linewidth}\raggedright
Excess reactivity
\end{minipage} & \begin{minipage}[b]{\linewidth}\raggedright
2430 \ensuremath{\pm} 14
\end{minipage} & \begin{minipage}[b]{\linewidth}\raggedright
2457 (Purgato, 2014)
\end{minipage} \\
\begin{minipage}[b]{\linewidth}\raggedright
Configuration 1
\end{minipage} & \begin{minipage}[b]{\linewidth}\raggedright
-3 \ensuremath{\pm} 9
\end{minipage} & \begin{minipage}[b]{\linewidth}\raggedright
9 \ensuremath{\pm} 2 (Santos et. al., 2008b)
\end{minipage} \\
\begin{minipage}[b]{\linewidth}\raggedright
Configuration 2
\end{minipage} & \begin{minipage}[b]{\linewidth}\raggedright
7 \ensuremath{\pm} 14
\end{minipage} & \begin{minipage}[b]{\linewidth}\raggedright
1 \ensuremath{\pm} 2 (Santos et. al., 2008b)
\end{minipage} \\
\begin{minipage}[b]{\linewidth}\raggedright
Fraction of delayed neutrons
\end{minipage} & \begin{minipage}[b]{\linewidth}\raggedright
743 \ensuremath{\pm} 10
\end{minipage} & \begin{minipage}[b]{\linewidth}\raggedright
750 \ensuremath{\pm} 4 (Santos and Diniz, 2014)
\end{minipage} \\
\midrule\noalign{}
\endhead
\bottomrule\noalign{}
\endlastfoot
\end{longtable}

The data in table 5 show good agreement with the calculations obtained by computational modeling via MCNP, showing about 1.11\% difference in excess reactivity, 12 pcm of difference for configuration 1, 6 pcm of difference for configuration 2 and less than 1\% for the measurement of the fraction of delayed neutrons. The differences in uncertainty and values seen in table 5 can be attributed to the author\textquotesingle s use of a personal computer, reducing the number of stories in the monte carlo simulations for a more precise calculation.

Finally, as the experiment will deal with the measurement of the integral reactivity of the control bar, we compare the measurements made of the integral reactivity of the control bar via the stable period method with the reactivity calculated with the MCNP 6.2, in the case of the MCNP the reactivity is estimated based on the comparison with the fully inserted bar case and the comparison can be seen in figure 3:

\includegraphics[width=4.50379in,height=3.44896in]{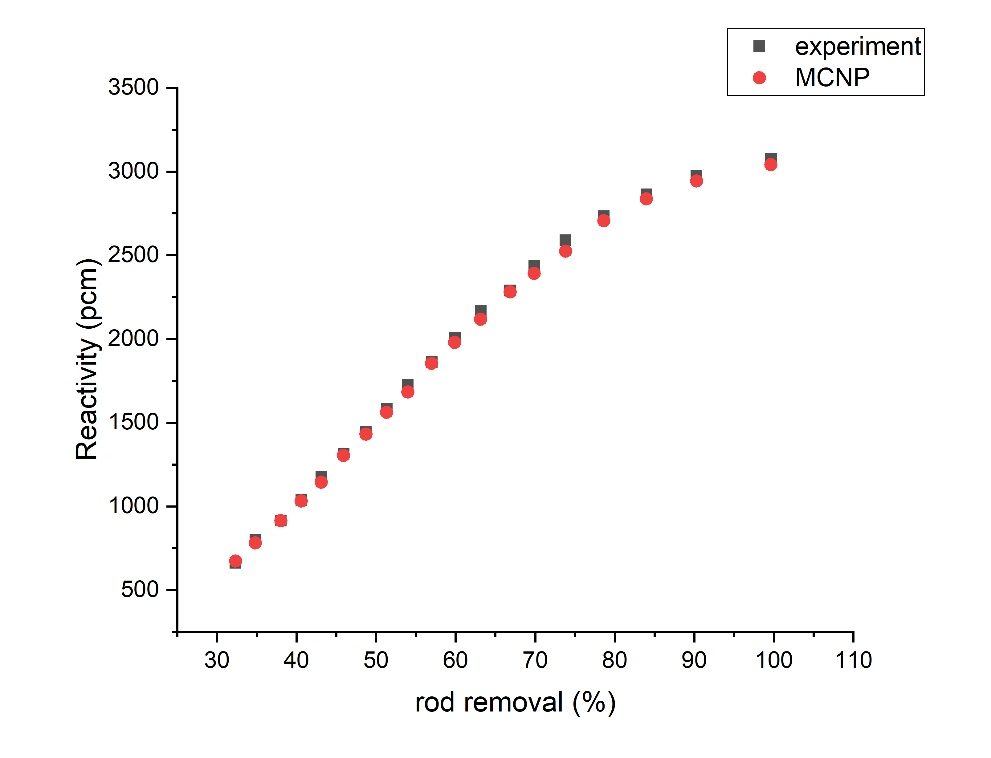}

Figure 3: Control bank Calibration.

Taking the results from table 5 and figure 3, it is possible to conclude that the model agrees significantly with the calculations of the reactor and thus we can extrapolate that it can be used to evaluate the rod-drop experiment. The reason why the value of the effective fraction of delayed neutrons is evaluated is that will be necessary to use the prompt jump approximation in the case of the rod-drop model.

4.2. Rod drop experiment

The reactivity measurement using the rod-drop technique consisted of criticalizing the reactor by leaving the control rod B 32\% removed and the control rod A 100\% removed, then letting the control rod A fall with gravity, the reactivity value was measured with the a rhetimeter, which reads the detectors that are described in figure 2 and the measurement obtained can be seen below in figures 4 and 5:

\includegraphics[width=4.57004in,height=3.49458in]{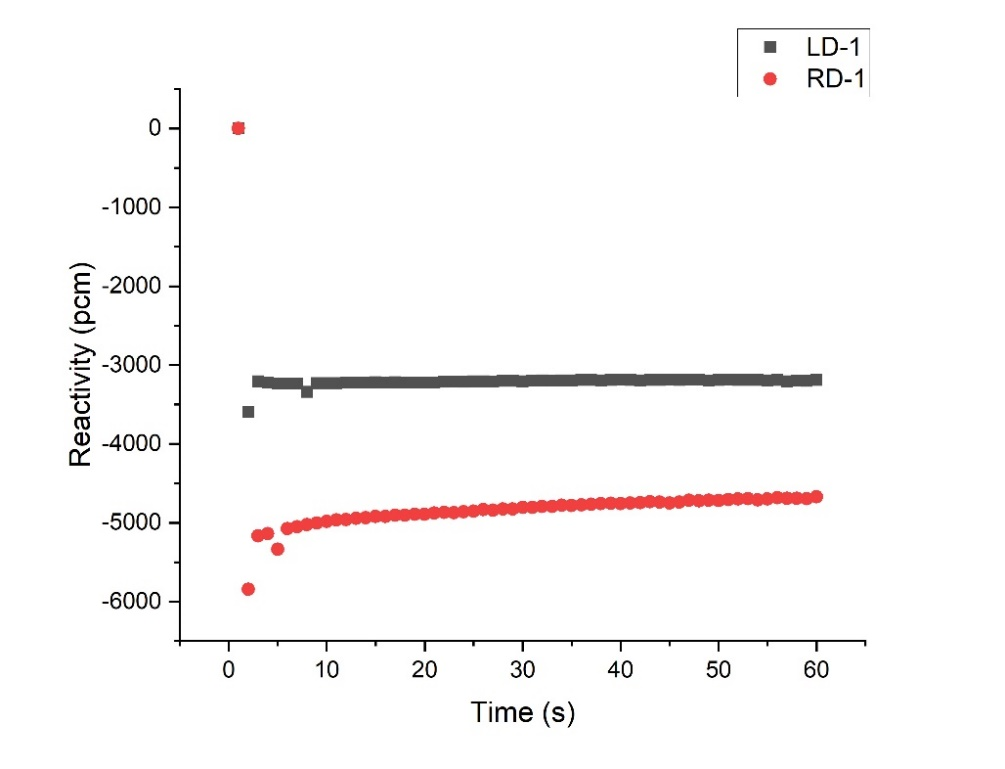}

Figure 4: Reactivity measurement using the rod drop method measured in the RD-1 and LD-1 detectors.

\includegraphics[width=4.94877in,height=3.85483in]{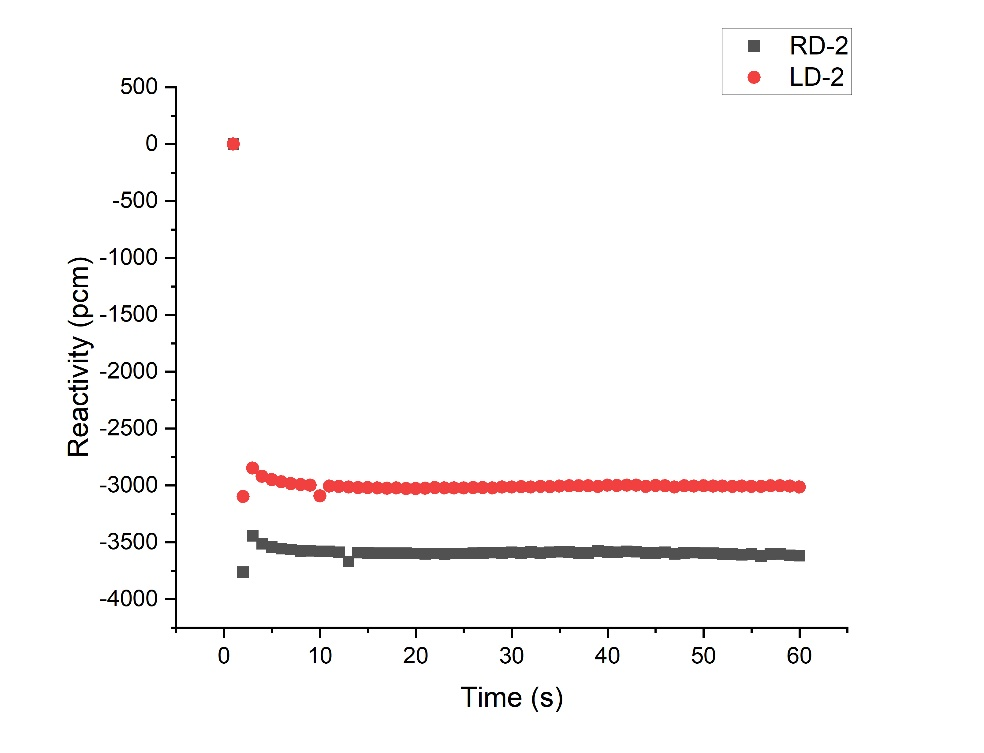}

Figure 5: Reactivity measurement using the rod drop method measured in the RD-2 and LD-2 detectors.

Table 6: Measuring reactivity by means of the 60 s after dropping control bar A.

\begin{longtable}[]{@{}
  >{\raggedright\arraybackslash}p{(\columnwidth - 2\tabcolsep) * \real{0.3241}}
  >{\raggedright\arraybackslash}p{(\columnwidth - 2\tabcolsep) * \real{0.6759}}@{}}
\toprule\noalign{}
\begin{minipage}[b]{\linewidth}\raggedright
\textbf{Detector}
\end{minipage} & \begin{minipage}[b]{\linewidth}\raggedright
\textbf{Reactivity measurement (pcm)}
\end{minipage} \\
\begin{minipage}[b]{\linewidth}\raggedright
LD-1
\end{minipage} & \begin{minipage}[b]{\linewidth}\raggedright
-3214 \ensuremath{\pm} 24
\end{minipage} \\
\begin{minipage}[b]{\linewidth}\raggedright
RD-1
\end{minipage} & \begin{minipage}[b]{\linewidth}\raggedright
-4850 \ensuremath{\pm} 126
\end{minipage} \\
\begin{minipage}[b]{\linewidth}\raggedright
LD-2
\end{minipage} & \begin{minipage}[b]{\linewidth}\raggedright
-3592 \ensuremath{\pm} 17
\end{minipage} \\
\begin{minipage}[b]{\linewidth}\raggedright
RD-2
\end{minipage} & \begin{minipage}[b]{\linewidth}\raggedright
-3007 \ensuremath{\pm} 16
\end{minipage} \\
\midrule\noalign{}
\endhead
\bottomrule\noalign{}
\endlastfoot
\end{longtable}

From the figures 2 and 3 is possible  to note that the transient drops quickly (\textasciitilde10 s), this is due to 5 of the 6 groups of precursors having a half-life of less than 10 s, after which time the measurement becomes approximately constant. However,  there is a great distinction of values according to the position of the detector in relation to the control bank A dropped in the experiment, this is due to the distortion of the neutron flux during the measurement, which ends up being deformed by the action of the highly neutron absorber introduced quickly into the system. Due to this reason, the results of the reactivity measured in the neutron detectors lose their proportionality and the experiment no longer represents the reactivity of the core, requiring a calibration to recover this proportionality. To analyze the results, it is necessary to better evaluate this effect.

3.3. Quantifying the neutron flux redistribution (form factor)

For a more detailed and accurate analysis of the measurement of the integral reactivity value of the control bank to be obtained, we must analyze the neutron flux redistribution on the detectors. The basic principle arising from point kinetics is that the neutron flux variation in the detector is proportional to the neutron flux in the reactor core.

The evaluation of the reactivity in each detector will be determined using the Source Jerk technique, since when we insert a control bank the source term of prompt neutrons is removed, leaving only the precursors of delayed neutrons, which would be similar to source removal of the Source Jerk method.

To simulate this behavior in the MCNP code one must consider a neutron source of very small intensity in the center of the reactor whose sole purpose is to provide a neutron intensity to the simulation. This treatment with the source in the center provides a very big disadvantage in the simulation, which is the fact that in the critical position of the reactor, it becomes supercritical, reaching the state of criticality ahead of time. From a data analysis point of view, as the source will be present before and after the insertion of the control bar, it will not affect the inserted negative reactivity value. However, the MCNP code cannot reach a convergence of values in the supercritical state and thus we will only have measurements in later states. This difficulty demands a linear extrapolation to obtain the first point of the simulation.

Assuming the relationship between the average neutron flux in the fuel (\(\varphi_{F}\)) and the average neutron flux in the detector (\(\varphi_{D}\)) at the position of the control bar x, it is possible to write that:

\(f(x) = \frac{\varphi_{D}(x)}{\varphi_{F}(x)}.\) (19)

In ideal behavior, \emph{f(x)} is expected to be a constant, however, due to flux distortion this value will vary. Thus, it is possible to quantify the shadowing effect \emph{F(x)} as a ratio between the value at position x with respect to the value at the initial position x = 0:

\(F(x) = \frac{f(x)}{f(0)}.\) (20)

From this analysis, the following quantifications of the form factor on the detectors can be seen in figures 6 and 7:

\includegraphics[width=4.45175in,height=3.40911in]{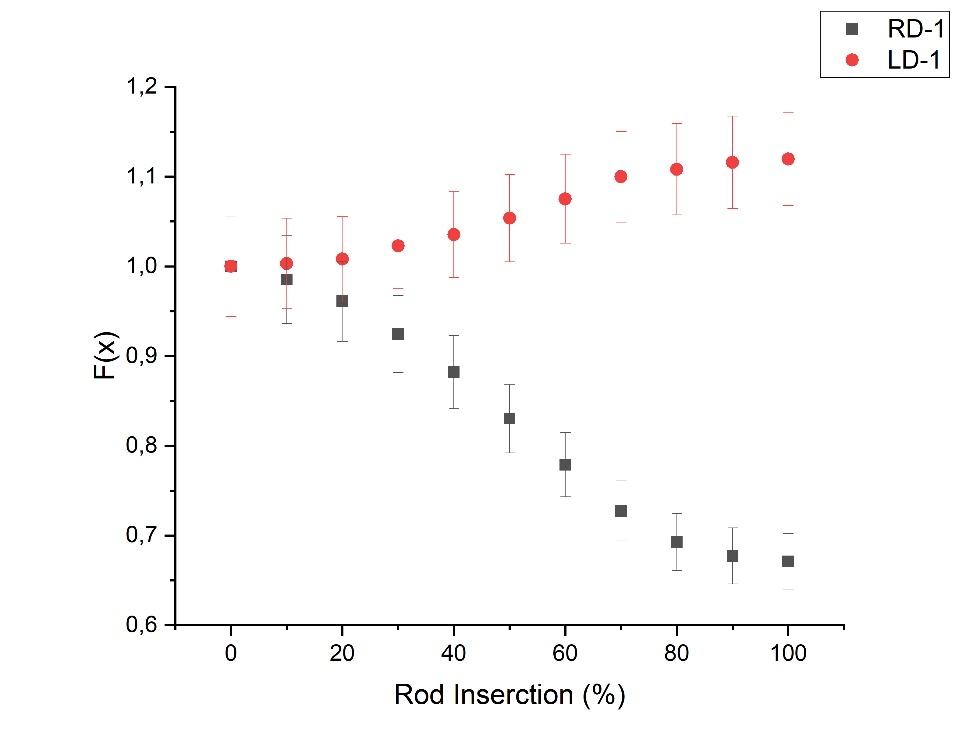}

Figure 6: Quantification of the form factor as a function of the insertion of control bar A for detectors RD-1 and LD-1.

\includegraphics[width=4.39872in,height=3.36851in]{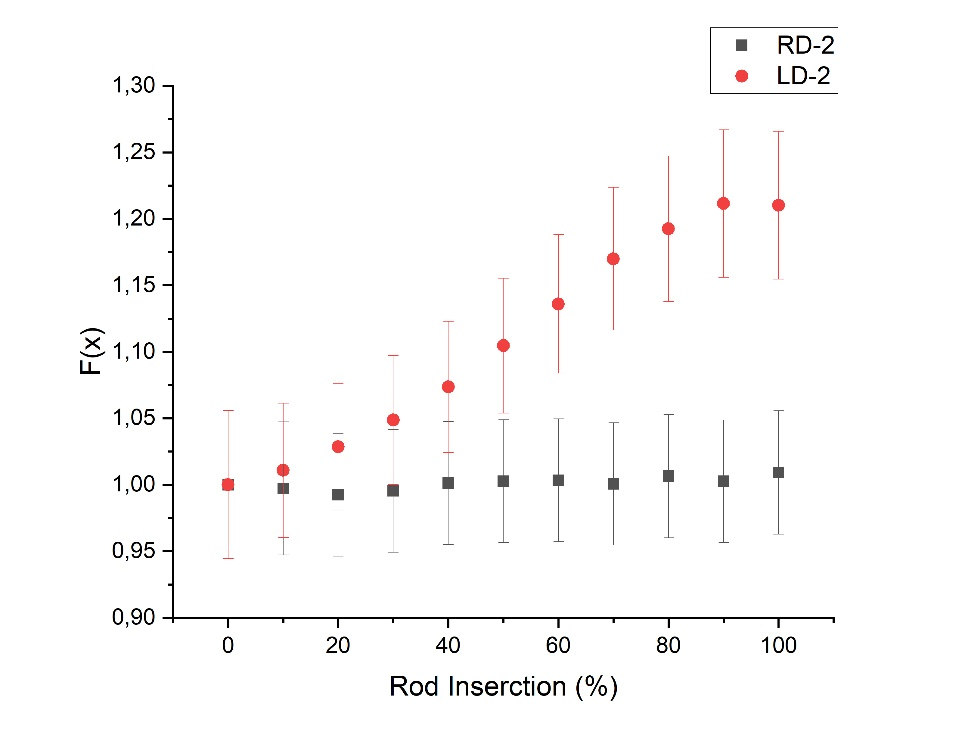}

Figure 7: Quantification of the form factor as a function of the insertion of control bar A for the RD-2 and LD-2 detectors.

4.4. Calibration of the reactivity measured by the detector

From the values obtained in figures 6 and 7 for the detectors, it is possible to apply corrections to the reactivity, which is nothing more than a calibration of the instrument to take into account the form factor in the measurement of the rod worth reactivity. For this, the reactivity value is needed in each simulated detector, and the Source Jerk technique was used. For value comparison purposes, as the result of the Source Jerk predicts the reactivity immediately after the fall of control rod A, the experimental value is taken just after the fall (first point) and compared with the value obtained via calculation of the MCNP shown in table 7 below:

Table 7: Measurement of reactivity in uncalibrated detectors.

\begin{longtable}[]{@{}
  >{\raggedright\arraybackslash}p{(\columnwidth - 4\tabcolsep) * \real{0.2516}}
  >{\raggedright\arraybackslash}p{(\columnwidth - 4\tabcolsep) * \real{0.3473}}
  >{\raggedright\arraybackslash}p{(\columnwidth - 4\tabcolsep) * \real{0.4012}}@{}}
\toprule\noalign{}
\multirow{3}{=}{\begin{minipage}[b]{\linewidth}\raggedright
\end{minipage}} & \multicolumn{2}{>{\raggedright\arraybackslash}p{(\columnwidth - 4\tabcolsep) * \real{0.7484} + 2\tabcolsep}@{}}{%
\begin{minipage}[b]{\linewidth}\raggedright
\textbf{Uncalibrated detectors}
\end{minipage}} \\
& \multicolumn{2}{>{\raggedright\arraybackslash}p{(\columnwidth - 4\tabcolsep) * \real{0.7484} + 2\tabcolsep}@{}}{%
\begin{minipage}[b]{\linewidth}\raggedright
Reactivity (pcm)
\end{minipage}} \\
& \begin{minipage}[b]{\linewidth}\raggedright
MCNP
\end{minipage} & \begin{minipage}[b]{\linewidth}\raggedright
Experiment
\end{minipage} \\
\begin{minipage}[b]{\linewidth}\raggedright
RD-2
\end{minipage} & \begin{minipage}[b]{\linewidth}\raggedright
-3968 \ensuremath{\pm} 175
\end{minipage} & \begin{minipage}[b]{\linewidth}\raggedright
-3762 \ensuremath{\pm} 107
\end{minipage} \\
\begin{minipage}[b]{\linewidth}\raggedright
LD-2
\end{minipage} & \begin{minipage}[b]{\linewidth}\raggedright
-3186 \ensuremath{\pm} 145
\end{minipage} & \begin{minipage}[b]{\linewidth}\raggedright
-3097 \ensuremath{\pm} 82
\end{minipage} \\
\begin{minipage}[b]{\linewidth}\raggedright
LD-1
\end{minipage} & \begin{minipage}[b]{\linewidth}\raggedright
-3504 \ensuremath{\pm} 158
\end{minipage} & \begin{minipage}[b]{\linewidth}\raggedright
-3593 \ensuremath{\pm} 136
\end{minipage} \\
\begin{minipage}[b]{\linewidth}\raggedright
RD-1
\end{minipage} & \begin{minipage}[b]{\linewidth}\raggedright
-6344 \ensuremath{\pm} 263
\end{minipage} & \begin{minipage}[b]{\linewidth}\raggedright
-5845 \ensuremath{\pm} 274
\end{minipage} \\
\begin{minipage}[b]{\linewidth}\raggedright
K-code
\end{minipage} & \begin{minipage}[b]{\linewidth}\raggedright
-4049 \ensuremath{\pm} 4
\end{minipage} & \begin{minipage}[b]{\linewidth}\raggedright
-
\end{minipage} \\
\midrule\noalign{}
\endhead
\bottomrule\noalign{}
\endlastfoot
\end{longtable}

From table 7 it is possible to notice that the reactivity value measured by the detector less affected by the shadowing effect (RD-2) was much higher than the reactivity value obtained by the stable period experiment obtained in figure 3 (3041 \ensuremath{\pm} 5 pcm). To determine the reason for this discrepancy, the k-code card of the MCNP code was used, which provided a negative reactivity value equal to 4049 \ensuremath{\pm} 4 pcm, which demonstrates that in addition to the effect of flux redistribution caused by the fall of the control bank, there is also the effect of increased negative reactivity caused by the presence of another partially inserted control bank nearby to criticalize the reactor, called shadowing effect. In the stable period experiment, as one rod is removed while another is inserted to make the reactor critical again, this effect is compensated, which does not occur in the rod drop experiment. The reactivities considered in Table 7 are higher because they were measured during the highest transient state (t \textless{} 10 s). In this sense, the measurement variance in this time interval was used as uncertainty. The higher uncertainty values derived from the MCNP code were due to the neutron histories used in the calculations and computational limitation of the available personal computer.

At this point an additional correction regarding the inserted bar is proposed, which is written as follows:

\(\rho_{c} = \frac{n - n_{0}f(100)}{n} \bullet \varepsilon \bullet \beta.\) (21)

or otherwise:

\(\rho_{c} = f(100) \bullet \rho \bullet \varepsilon - \left( f(100) - 1 \right) \bullet \varepsilon \bullet \beta,\) (22)

where \(\rho_{c}\) is the calibrated reactivity value, f(100) the form factor for the inserted 100\% bar, \ensuremath{\varepsilon} the ratio between the actual value of the control bank and the value shadowing by the presence of another control bank, and \ensuremath{\rho} the measured value. The results obtained by applying the correction proposed in equation (22) to the data in table 7 can be seen in table 8 below:

Table 8: Measurement of reactivity in calibrated detectors

\begin{longtable}[]{@{}
  >{\raggedright\arraybackslash}p{(\columnwidth - 8\tabcolsep) * \real{0.1473}}
  >{\raggedright\arraybackslash}p{(\columnwidth - 8\tabcolsep) * \real{0.3095}}
  >{\raggedright\arraybackslash}p{(\columnwidth - 8\tabcolsep) * \real{0.1306}}
  >{\raggedright\arraybackslash}p{(\columnwidth - 8\tabcolsep) * \real{0.2827}}
  >{\raggedright\arraybackslash}p{(\columnwidth - 8\tabcolsep) * \real{0.1298}}@{}}
\toprule\noalign{}
\multirow{3}{=}{\begin{minipage}[b]{\linewidth}\raggedright
\end{minipage}} & \multicolumn{4}{>{\raggedright\arraybackslash}p{(\columnwidth - 8\tabcolsep) * \real{0.8527} + 6\tabcolsep}@{}}{%
\begin{minipage}[b]{\linewidth}\raggedright
\textbf{Calibrated detectors}
\end{minipage}} \\
& \multicolumn{2}{>{\raggedright\arraybackslash}p{(\columnwidth - 8\tabcolsep) * \real{0.4401} + 2\tabcolsep}}{%
\begin{minipage}[b]{\linewidth}\raggedright
MCNP
\end{minipage}} & \multicolumn{2}{>{\raggedright\arraybackslash}p{(\columnwidth - 8\tabcolsep) * \real{0.4125} + 2\tabcolsep}@{}}{%
\begin{minipage}[b]{\linewidth}\raggedright
Experiment
\end{minipage}} \\
& \begin{minipage}[b]{\linewidth}\raggedright
Reactivity (pcm)
\end{minipage} & \begin{minipage}[b]{\linewidth}\raggedright
\(\mathrm{\Delta}\rho\) (\%)
\end{minipage} & \begin{minipage}[b]{\linewidth}\raggedright
Reactivity (pcm)
\end{minipage} & \begin{minipage}[b]{\linewidth}\raggedright
\(\mathrm{\Delta}\rho\) (\%)
\end{minipage} \\
\begin{minipage}[b]{\linewidth}\raggedright
RD-2
\end{minipage} & \begin{minipage}[b]{\linewidth}\raggedright
-3030 \ensuremath{\pm} 211
\end{minipage} & \begin{minipage}[b]{\linewidth}\raggedright
0.35
\end{minipage} & \begin{minipage}[b]{\linewidth}\raggedright
-2877 \ensuremath{\pm} 194
\end{minipage} & \begin{minipage}[b]{\linewidth}\raggedright
6.40
\end{minipage} \\
\begin{minipage}[b]{\linewidth}\raggedright
LD-2
\end{minipage} & \begin{minipage}[b]{\linewidth}\raggedright
-3043 \ensuremath{\pm} 211
\end{minipage} & \begin{minipage}[b]{\linewidth}\raggedright
0.06
\end{minipage} & \begin{minipage}[b]{\linewidth}\raggedright
-2727 \ensuremath{\pm} 188
\end{minipage} & \begin{minipage}[b]{\linewidth}\raggedright
11.33
\end{minipage} \\
\begin{minipage}[b]{\linewidth}\raggedright
LD-1
\end{minipage} & \begin{minipage}[b]{\linewidth}\raggedright
-3036 \ensuremath{\pm} 212
\end{minipage} & \begin{minipage}[b]{\linewidth}\raggedright
0.17
\end{minipage} & \begin{minipage}[b]{\linewidth}\raggedright
-2986 \ensuremath{\pm} 218
\end{minipage} & \begin{minipage}[b]{\linewidth}\raggedright
2.87
\end{minipage} \\
\begin{minipage}[b]{\linewidth}\raggedright
RD-1
\end{minipage} & \begin{minipage}[b]{\linewidth}\raggedright
-3056 \ensuremath{\pm} 213
\end{minipage} & \begin{minipage}[b]{\linewidth}\raggedright
0.50
\end{minipage} & \begin{minipage}[b]{\linewidth}\raggedright
-3163 \ensuremath{\pm} 227
\end{minipage} & \begin{minipage}[b]{\linewidth}\raggedright
2.95
\end{minipage} \\
\midrule\noalign{}
\endhead
\bottomrule\noalign{}
\endlastfoot
\end{longtable}

In calibrating the experimental data, the effect of neutron flux redistrubtion on the detectors and shadowing effect of the control banks obtained through the calculation with the MCNP code was considered. The values obtained shown good agreement with the proposed correction and maximum percentage deviations of 0.50\% theoretical and 11.33\% experimental.

\begin{enumerate}
\def\labelenumi{\arabic{enumi}.}
\setcounter{enumi}{4}
\item
  CONCLUSION
\end{enumerate}

The validation presented in the first part of the results demonstrates that the model used in the simulation of the IPEN/MB-01 reactor has a maximum difference of theoretical results of 13 pcm when compared with the experimental results of the references.

The measurements of the rod drop experiment demonstrate the shadowing effect and flux redistribution effect, which would be the loss of proportionality of the reactivity measured in the detectors in relation to the reactivity of the nucleus. Although the MCNP code does not have a routine for simulating transient results, it was possible to detect the shading effect with a certain approximation of the state close, temporally, to the fall of the control bank in the experiment.

Simulation with the MCNP Kcode routine demonstrates that the negative reactivity inserted by a single control bank has its negative value magnified by the presence of another partially inserted bank by about 1000 pcm. Thus, it is noted that the calibration must both take into account the value of the shadowing effect due the control bank parcialy inserted as the neutron flux distortion.

From the quantification of the shadowing effect, defined in the results analysis section, a calibration for the neutron flux was proposed. As an application, the source jerk method was used to correct the reactivity value obtained by the detector. The maximum theoretical and experimental differences were 0.50\% and 11.33\%.

For future studies, it is expected to use or develop a neutron transport code that uses the temporal variation of the neutron flux so that the transient behavior of the neutron flux after the fall of the control bank can be more adequately represented, with this it would be possible to use the region with the smallest reactivity transient as a comparison of results, that is, t \textgreater{} 10 s. Despite the limitations of the study with quasi-static codes such as MCNP, there is good agreement between theoretical and experimental results in the rod drop experiment.

\textbf{ACKNOWLEDGMENTS}

This research project is supported by the following Brazilian institutions: Brazilian Council for Scientific and Technological Development (CNPq), Brazilian Nuclear Energy Commission (CNEN), and Research Support Foundation of the State of Rio de Janeiro (FAPERJ).

\textbf{REFERENCES}

Asuku, A. et. al., 2015. Application of positive period method in the calibration and determination of integral worth of MNSR control rod. Int. J. Nucl. Energy Sci. Technol. 9 (4), 319--332.

https://doi.org/10.1504/IJNEST.2015.075485

Buoni, F. B., Experience with the use of the rod-drop method of rod calibration at the ORR and IITR, 1963, Oak Ridge National Laboratory.

Duderstadt, J. J., Hamilton, L. J. Nuclear Reactor Analysis, Ed Joh Wiley and Sons, Nova York (1976).

Endo, T. et. al., 2011. Detected-neutron Multiplication factor measured by neutron source multiplication method, Annals of Nuclear Energy, vol. 38, 2417-2427.

https://doi.org/10.1016/j.anucene.2011.01.007

Fadaei, A. H. and Setayeshi, S., 2009. Control rod worth calculation for VVER-1000 nuclear reactor using WIMS and CITATION codes, Progress in Nuclear Energy 51, 184-191.

https://doi.org/10.1016/j.pnucene.2008.03.003

Gauthier, J. C. et. al., 2017. Measurement and Predictions of Control Rod Worth, Nuclear Science and Engineering 106, 18-29.

https://doi.org/10.13182/NSE90-02

Holt, L. et. al., 2016. Investigation of feedback on neutron kinetics and thermal hydraulics from detailed online fuel behavior modeling during a boron dilution transient in a PWR with the two-way coupled code system DYN3D-TRANSURANUS, Nuclear Engineering and Design 297, 32-43.

https://doi.org/10.1016/j.nucengdes.2015.11.005

Kaviani, P., Mokhtar, J. and Mirvakili, S. M., 2020. Integral form of the control rod calibration curve in the new core configuration of HWZPR using rod insertion method. Progress in Nuclear Energy, 125.

https://doi.org/10.1016/j.pnucene.2020.103375

Kloosterman, J.L., Rugama, Y. and M. Szieberth, 2002. MEASUREMENT AND CALCULATION OF CONTROL ROD WORTHS IN MASURCA, PHYSOR, Seuoul Korea, 7-10, October.

MARGUET, S. The physics of nuclear reactors. Springer, 2018.

Mele, I., Ravnik, M. and Trkov, A., 2017. TRIGA Mark II Benchmark Experiment, Part I: Steady-State Operation, Nuclear Technology 105.

https://doi.org/10.13182/NT94-A34909

Merljak, V., Kromar, M. and Trkov, A., 2018. Rod insertion method analysis - A methodology update and comparison to boron dilution method. Annals of Nuclear Energy 113, 96-104.

https://doi.org/ 10.1016/j.anucene.2017.11.020

Moreira J. and Lee, J. C., 1984. Space-Time Analysis of Reactor Control Rod Worth Measurements. Nuclear Science and Engineering 86, 91-105.

https://doi.org/10.13182/NSE84-1

Moreira, J., Real time reactivity measurements - \textquotesingle reatimetro\textquotesingle, 1986, Rio de Janeiro, RJ - Brasil.

Nelson, G.W. and Doane, H.J., 1990. Rapid and Accurate Control Rod Calibration Measurement and Analysis. No. INIS-US--09N0043.

Pinto, L. N., 2012. Reactivity Effects Experiments at IPEN/MB-01 Nuclear Reactor. Dissertação de Mestrado. Master\textquotesingle s thesis, IPEN/USP.

Purgato, R. P., 2014. Determination of buckling and probability of leakage of neutron in the IPEN/MB-01 reactor in cylindrical configuration. Master's thesis -- IPEN/USP.

Pyeon, C. H. et al., 2017. Benchmarks of subcriticality in accelerator-driven system at Kyoto University Critical Assembly, Nuclear engineering and technology, 49(6), 1234-1239. https://doi.org/10.1016/j.net.2017.06.012

Roy, T. et. al., 2017. Pulsed Neutron Source Measurements in the BRAHMMA Accelerator-Driven Subcritical System, Nuclear Science and Engineering 184, 584-590. https://doi.org/10.13182/NSE16-81

Ryves, T. B. and Scott, M. C., 1962. Subcritical reactivity measurement by a source jerk method, Journal of Nuclear Energy, vol. 16, 455-463.

https://doi.org/10.1016/0368-3230(62)90198-2

Santos, A. et. al, 2006. Isothermal Experiments of The IPEN/MB-01 Reactor, International Handbook of Evaluated Reactor Physics Benchmark Experiments, Light Water Moderated Reactor -- LWR, NEA/NSC/DOC (2006), p. 58, 2006.

Santos, A. et. al., 2008a. Reactor Physics Experiments in the IPEN/MB-01 Research Reactor Facility, March, Paris: Blair Briggs.

Santos, A. et. al., 2008b. Critical loading configurations of the IPEN/MB-01 reactor with a heavy SS-3 reflector LEU-COMP-THERM-043. In. International Handbook of Evaluated Criticality Safety Benchmark Experiments. Paris: NEA/NSC, 2007.

Santos, A. et. Al, 2013. Validation of neutronic models and calculation systems by means of experimental results in the IPEN/MB-01 reactor. Annals of Nuclear Energy 60, 51-63.

https://doi.org/10.1016/j.anucene.2013.04.022

Santos, A. and Diniz, R., 2014. The Evaluation of the Effective Kinetic Parameters and Reactivity of the IPEN/MB-01 Reactor for the International Reactor Physics Experiment Evaluation Project, Nuclear Science and Engineering, 178:4, 459-478.

https://doi.org/10.13182/NSE14-10

Sembiring, T. M. et. al, 2021. Analysis of the excess reactivity and control rod worth of RSG-GAS equilibrium silicide core using Continuous-Energy Monte Carlo Serpent2 code, Annals of Nuclear Energy.

https://doi.org/10.1016/j.anucene.2020.108107

Son, N. A. et. A., 2017. Control Rod Calibration and Worth Calculation for Optimized Power Reactor 1000 (OPR-1000) Using Core Simulator OPR1000. World Journal of Nuclear Science and Technology, 7, 15-23.

https://doi.org/10.4236/wjnst.2017.71002

Thilagam, L. et. al., 2021. Simulations for worth determination of CPSARs in VVER-1000 reactor using EXCEL and TRIHEX-FA code system. Annals of Nuclear Energy, 151.

https://doi.org/10.1016/j.anucene.2020.107921

Trkov, A. et. al, 1995. Application of the rod-insertion method for control rod worth measurements in research reactors Kerntechnik, vol. 60, no. 5-6, 255-261.

https://doi.org/10.1515/kern-1995-605-620

Varvayanni, M., Savva, P. and Catsaros, N., 2009. Control rod worth calculations using deterministic and stochastic methods, Annals of Nuclear Energy 36, 1718-1725.

https://doi.org/10.1016/j.anucene.2009.09.003

\end{document}